\documentclass[prl,twocolumn,showpacs,amsmath,amssymb,superscriptaddress]{revtex4-1}
\usepackage{graphicx}
\usepackage{breqn}
\usepackage{xcolor}
\usepackage{color}
\usepackage{booktabs}
\usepackage{float}
\usepackage{longtable}
\usepackage{epsfig}
\usepackage{bm}
\usepackage{dcolumn}
\usepackage{lipsum, babel}
\usepackage{soul}
\usepackage{cancel}

\usepackage{rotating} 

\usepackage[normalem]{ulem}
\usepackage{epstopdf}

\usepackage{multirow}
\usepackage{booktabs, makecell}

\makeatletter
\let\cat@comma@active\@empty
\makeatother
\begin{document}

\title{Atomic altermagnetism}

\author{Rodrigo Jaeschke-Ubiergo}
\affiliation{Institut f\"ur Physik, Johannes Gutenberg Universit\"at Mainz, D-55099 Mainz, Germany}

\author{Venkata-Krishna Bharadwaj}
\affiliation{Institut f\"ur Physik, Johannes Gutenberg Universit\"at Mainz, D-55099 Mainz, Germany}

\author{Warlley Campos}
\affiliation{Max Planck Institute for the Physics of Complex Systems, N\"othnitzer Str. 38, 01187 Dresden, Germany}

\author{Ricardo Zarzuela}
\affiliation{Institut f\"ur Physik, Johannes Gutenberg Universit\"at Mainz, D-55099 Mainz, Germany}

\author{Nikolaos Biniskos}
\affiliation{Charles University, Faculty of Mathematics and Physics, Department of Condensed Matter Physics, Ke Karlovu 5, 121 16, Praha, Czech Republic}

\author{Rafael M. Fernandes}
\affiliation{Department of Physics, The Grainger College of Engineering, University of Illinois Urbana-Champaign, Urbana, IL 61801, USA}
\affiliation{Anthony J. Leggett Institute for Condensed Matter Theory, The Grainger College of Engineering, University of Illinois Urbana-Champaign, Urbana, IL 61801, USA}

\author{Tomas~Jungwirth}
\affiliation{Institute of Physics, Czech Academy of Sciences, Cukrovarnick\'a 10, 162 00, Praha 6, Czech Republic}
\affiliation{School of Physics and Astronomy, University of Nottingham, NG7 2RD, Nottingham, United Kingdom}

\author{Jairo Sinova}
\affiliation{Institut f\"ur Physik, Johannes Gutenberg Universit\"at Mainz, D-55099 Mainz, Germany}

\author{Libor \v{S}mejkal}
\affiliation{Max Planck Institute for the Physics of Complex Systems, N\"othnitzer Str. 38, 01187 Dresden, Germany}
\affiliation{Max Planck Institute for Chemical Physics of Solids, N\"othnitzer Str. 40, 01187 Dresden, Germany}
\affiliation{Institut f\"ur Physik, Johannes Gutenberg Universit\"at Mainz, D-55099 Mainz, Germany}
\affiliation{Institute of Physics, Czech Academy of Sciences, Cukrovarnick\'a 10, 162 00, Praha 6, Czech Republic}

\date{\today}

\newcommand{\rj}[1]{\textcolor{teal}{#1}} 

\newcommand{\bcoo}{Ba$_2$CaOsO$_6$}
\newcommand{\ucr}{UCr$_2$Si$_2$C}
\newcommand{\aoms}{A$_2$O$_3$M$_2$Se$_2$}
\newcommand{\lofs}{La$_2$O$_3$Fe$_2$Se$_2$}
\newcommand{\loms}{La$_2$O$_3$Mn$_2$Se$_2$}

\newcommand{\vso}{V$_2$Se$_2$O}
\newcommand{\vto}{V$_2$Te$_2$O}

\newcommand{\kvso}{KV$_2$Se$_2$O}
\newcommand{\mo}{MoO}

\begin{abstract}

Altermagnetism has  been recently verified experimentally {by} photoemission mapping of the spin order in momentum space in MnTe and CrSb, which feature two anisotropic sublattices with antiparallel magnetic dipole moments. 
In this work, we explicitly demonstrate the presence of an even-parity 
 ferroically ordered non-dipolar spin density on atomic sites, i.e. atomic altermagnetism, in MnTe, \kvso\ and \bcoo. We do so  through spin-symmetry analysis and partial-wave decomposition of the spin density obtained by first-principles calculations. In MnTe we show 
 a ferroically ordered $g$-wave form factor in the spin density around the Mn 
 site.
   In \kvso\ (and related Lieb lattice compounds), we show that there is a ferroically ordered $d$-wave form factor coexisting with the antiferroic magnetic dipoles in the V site, while the O site shows no dipole but a pure $d$-wave atomic spin density.
In the Mott-insulating 
\bcoo, as a key result, we 
  reveal a pure form of atomic altermagnetism -- absent of any dipolar sublattice order.
This highlights that the altermagnetic order can exist without a Néel vector formed by antiferroic dipole moments on an even number of crystal sublattices, underlining  its distinction 
from collinear Néel  antiferromagnetic order. Our calculations predict that  \kvso\ and \bcoo\ can exhibit giant spin-splitter angles of up to 42° and 26° respectively, thus demonstrating the possibility of large altermagnetic responses without requiring the staggered Néel order of local dipole moments.


\end{abstract}

\maketitle

\begin{figure*}[ht!]
	\centering
	\includegraphics[scale=0.23]{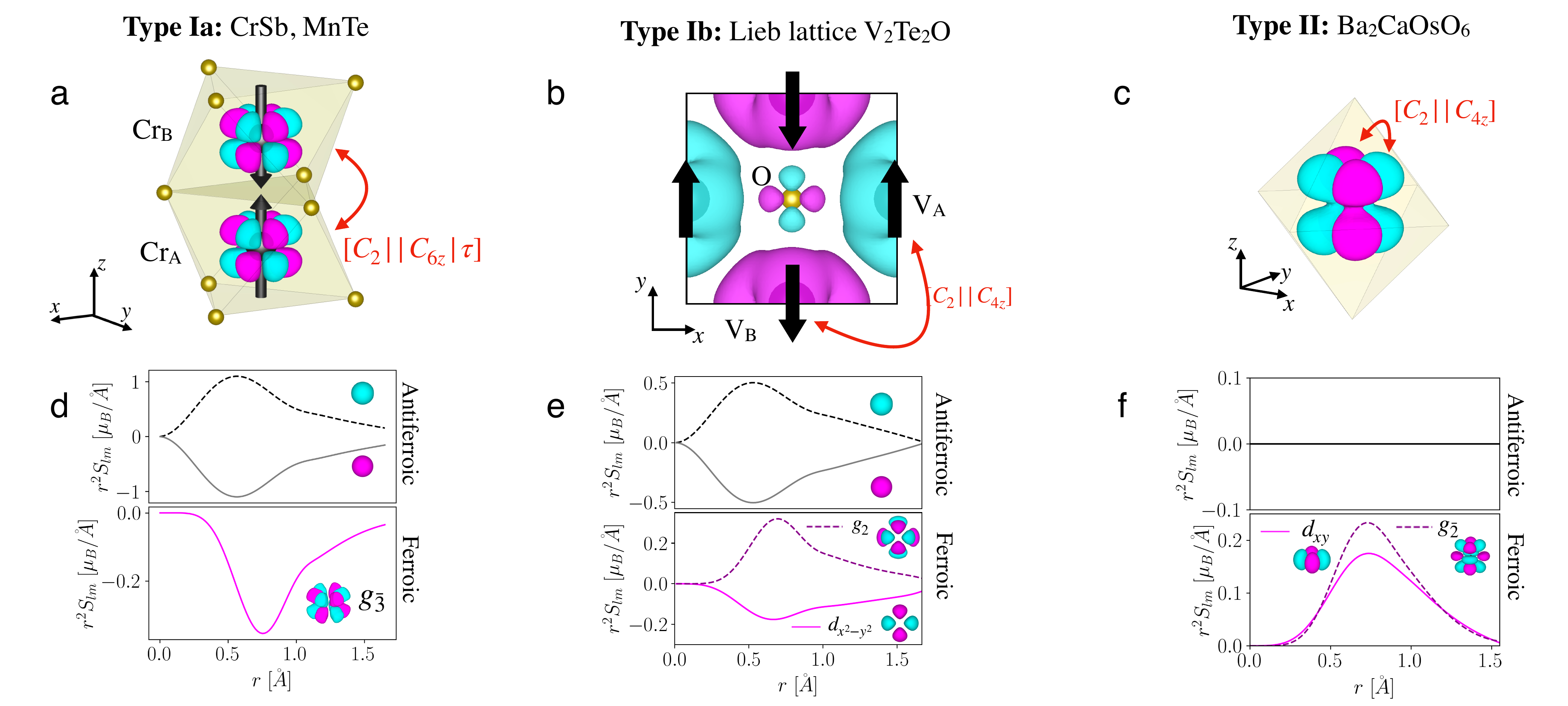}
	\caption{Partial-wave expansion of the spin density from non-relativistic DFT for CrSb, \vto, and \bcoo. Panels (d–f) show the radial profiles of the key partial-wave coefficients $S_{lm}(r)$: top panels display antiferroic $s$-wave (dipolar) components; bottom panels show ferroically ordered higher harmonics.
		(a) CrSb unit cell with antiferroic dipoles (black arrows) and ferroic $g_{\bar{3}} \equiv Y_{4,-3}$.
		(b) \vto\ unit cell and spin density iso-surface; V sites show antiferroic dipoles and ferroic $d_{x^2 - y^2}$ and $g_2 \equiv Y_{42}$ components; the O site hosts a pure $d_{x^2 - y^2}$ density.
		(c) \bcoo\ spin density iso-surface, showing no dipole but a pure $d$-wave spin density }
	\label{fig:Fig1}
\end{figure*}

Altermagnetism is a recently identified new type of magnetism characterized by spin order in both direct and momentum space with $d$-, $g$-, or $i$-wave symmetry  \cite{Smejkal2021a}. Altermagnetism was predicted by a rigorous classification and delimitation of collinear magnetic phases based on spin symmetries that involve pairs of generally distinct operations, acting on the lattice and spin degrees of freedom  \cite{Smejkal2021a}. These symmetries lead to unconventional spin densities \cite{Smejkal2020} and exchange fields  \cite{Smejkal2023}, which break the underlying lattice symmetry in an analogous way to what occurs in unconventional superfluid states \cite{Smejkal2022a,Jungwirth2024b}. Altermagnetism has been experimentally confirmed via photoemission spectroscopy and microscopy, in the binary-compounds MnTe and CrSb with ordering above room temperature \cite{Krempasky2024,Lee2024,Osumi2024,Reimers2024,Yang2024,Ding2024,Zeng2024,Li2024,Hariki2023, Amin2024}. More recently, two room-temperature metallic $d$-wave altermagnets, \kvso\ \cite{Jiang2025} and Rb$_{1-\delta}$V$_2$Te$_2$O \cite{Zhang2025a}, have been probed with neutron magnetic resonance, and spin resolved photo-emission measurements, respectively.

The discovery of the altermagnetic spin symmetry class also provided a unifying explanation of the  previous reports of unconventional time-reversal symmetry breaking (TRSB), such as the anomalous Hall effect \cite{Smejkal2020,Mazin2021}, spin currents  \cite{ Ahn2019,Naka2019,Gonzalez-Hernandez2021,Smejkal2022GMR} and magneto-optical effects  \cite{Samanta2020,Mazin2021}.
The unconventional electronic structure  
of altermagnets exhibits TRSB spin-split bands across the entire Brillouin Zone, except along 2 ($d$-wave), 4 ($g$-wave) or 6 ($i$-wave)
nodal surfaces  \cite{Smejkal2020,Mazin2021}, leading to various
spintronic effects  \cite{Smejkal2020,Mazin2021,Feng2022,Reichlova2024,Smejkal2022GMR,Bai2024,Naka2019,Shao2021}.
 
The  theoretically predicted and experimentally confirmed altermagnets are hitherto characterized by an even number of magnetic sublattices, whose alternating magnetic moments are accompanied by an alternating orientation of the atomic site environment in direct space  \cite{Smejkal2021a, Smejkal2022a,Guo2023b,Bai2024}.  This alternation of magnetic sites is commonly achieved by the anisotropic crystal environment of the magnetic sublattices  \cite{Smejkal2021a, Smejkal2020, Jaeschke-Ubiergo2023, Gomonay2024}, but can also be achieved by electronic correlations  \cite{Leeb2023, Duerrnagel2024, Giuli2025}.

Previous reports have suggested that altermagnetism is linked to ferroically ordered magnetic multipoles on the magnetic sublattices  \cite{Bhowal2024,Fernandes2023, McClarty2024, Verbeek2024, Schiff2025}. In this work we aim to quantify the altermagnetic direct-space order parameter \cite{Fernandes2023, McClarty2024}, by studying the partial-wave expansion of the  spin density obtained from Density Funtional Theory (DFT). We demonstrate the presence of a ferroically ordered $g$-wave form factor of the atomic spin density in CrSb and MnTe, which we consider here to be canonical altermagnets. In a series of compounds with a Lieb lattice structure \cite{Mazin2023a, Kaushal2024} (e.g., \kvso\  \cite{Jiang2025}), we find a ferroic $d$-wave spin density which is comparable in size with the antiferroic dipoles on V sites, and a pure $d$-wave spin density (no dipole) on the O site.

As a key result, in Ba$_2$CaOsO$_6$  \cite{Maharaj2020,Paramekanti2020,Voleti2021}, we propose a pure form of atomic altermagnetism, {i.e.}, a direct-space spin density without any dipolar moments at any atomic site but 
 with an on-site ferroically ordered $d$-wave form factor, thus enabling the realization of altermagnetism without the need to have the staggered Néel order of local dipole moments. We show that correlations and orbital ordering can generate $d$-wave altermagnetic spin density, without being promoted by particular crystal symmetry.  We also find large spin-splitter responses in the nonrelativistic limits of \kvso\ and \bcoo, demonstrating that pure atomic altermagnetism— in the case of \bcoo, without Néel order—can lead to large altermagnetic responses.

\begin{figure*}[ht]
	\centering
	\includegraphics[scale=0.21]{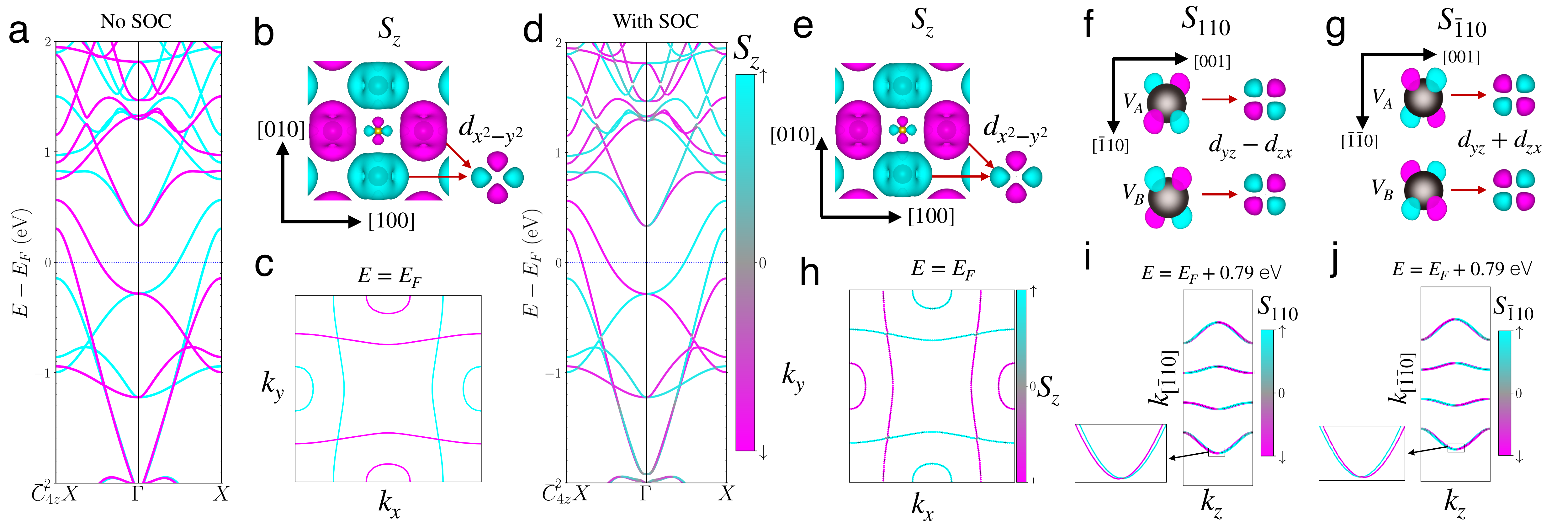}
	\caption{DFT results for \kvso\ without (a–c) and with (d–j) SOC.
		(a,d) Band structure with $S_z$ spin polarization shown in color scale.
		(b) Iso-surface of the non-relativistic spin density.
		(c) Energy iso-surface in the $k_z = 0$ plane.
		(e–g) Spin density iso-surfaces with SOC, projected along $[001]$, $[110]$, and $[\bar{1}10]$, respectively, showing ferroic partial-wave components: $d_{x^2 - y^2}$, $d_{zx} - d_{yz}$, and $d_{zx} + d_{yz}$, respectively.
		(h–j) Iso-energy lines in planes normal to $[001]$, $[110]$, and $[\bar{1}10]$, colored by spin polarization normal to each plane.}
	\label{fig:Fig2}
\end{figure*}

\textit{Partial-wave expansion of the spin density}|The spin density can be defined as $\mathbf{S}(\mathbf{r})=\frac{\hbar}{2}\sum_{\mathbf{k},n}f_{n\mathbf{k}} \psi_{n\mathbf{k}}^{\dagger}(\mathbf{r}) \boldsymbol{\sigma} \psi_{n\mathbf{k}}(\mathbf{r})$, where $ \psi_{n\mathbf{k}}(\mathbf{r})$ is the spinorial eigenfunction of wave-vector $\mathbf{k}$ and band index $n$, $f_{n\mathbf{k}}$ is the occupation number and $\boldsymbol{\sigma}$
are the Pauli matrices representing the spin degree of freedom. We define $\mathbf{S}_{\alpha}(\mathbf{r}) = \mathbf{S}(\mathbf{r}-\mathbf{r}_{\alpha})$ as the spin density with respect  to an atomic site $\mathbf{r}_{\alpha}$, and we further expand each spin component in partial waves as ${S}^i_{\alpha}(\mathbf{r}) = \sum_{lm} S^i_{\alpha,lm}(r) Y_{lm}(\theta, \phi)$ where $ Y_{lm}(\theta, \phi)$ are real spherical harmonics  \cite{Blanco1997} and $(r,\theta, \phi)$ denote the spherical coordinates of $\mathbf{r}-\mathbf{r}_{\alpha}$. The magnetic dipole approximation allows us to define an on-site magnetic dipole as the radial integral of the first term in the expansion $S^i_{\alpha,00}(r)$. The magnetic dipoles alone, however, are unable to capture the ferroic nature of altermagnetism  \cite{Bhowal2024,Fernandes2023}. 

\textit{Canonical altermagnets}|In a canonical altermagnet, such as CrSb (Type Ia, see Fig.~\ref{fig:Fig1}a,d), the magnetic dipoles align antiferroically, but the spin density around each magnetic site is anisotropic. For CrSb there is a ferroically ordered contribution with the symmetries of a $g_{\bar{3}}\equiv Y_{4,-{3}}$ harmonic (see Fig. \ref{fig:Fig1}d). Remarkably, unlike conventional multipolar orders which originate from spin-orbit coupling (SOC)  \cite{Santini2000,Zhao2016a}, the ferroic order in the above altermagnets is stabilized  by the antiferroic magnetic dipoles distributed in the crystal according to the altermagnetic spin group symmetry  \cite{Smejkal2021a}. The $g_{\bar{3}}$ harmonic has the same nodal structure as the spin splitting in the electronic band structure, which results from the spin group of CrSb containing operations such as $[C_{2}\vert\vert C_{6z}\vert\vec{\tau}]$ combining a two-fold spin rotation with a six-fold lattice rotation, and a half unit cell translation along the z-axis as marked in Fig.~1a. In the Supplementary Information (SI), we show similar analysis for MnTe.

\textit{Lieb lattice candidates}|Next we discuss a series of materials whose crystal structure can be understood as stacked layers of Lieb lattices \cite{Mazin2023a}. This list includes the quasi 2D monolayers of V$_2$A$_2$O (A=Se, Te) \cite{Lin2018, Qi2024}, and the bulk  structures \kvso\ \cite{Jiang2025}, \aoms\ (A=La, M=Mn, Fe)  \cite{Wei2025, Landsgesell2013, Free2011}  and \ucr\ \cite{Lemoine2018}.  

In Fig. \ref{fig:Fig1}b we show a top view of a Lieb lattice layer in \vto. We plot as well an iso-surface of the spin density obtained by DFT without SOC. This spin density follows the spin point group \textbf{ $^24/^1m^1m^2m$}, which classifies it as a $d$-wave altermagnet. Sites $A$ and $B$ are occupied by V atoms with in-plane antiparallel magnetic dipole moments. The radial profile of this antiferroic $s$-wave  spin density is shown in the top panel of Fig  \ref{fig:Fig1}e. At sites $A$ and  $B$ we identify also two higher order partial waves, that are ferroically aligned (same sign in both sublattices). These ferroic contributions are characterized by real spherical harmonics $d_{x^2-y^2}$ and $g_{2}\equiv Y_{4,2}$, both sharing the same $d$-wave structure in the $xy$ plane, with nodes along the diagonals. Notably, at the $O$ site, which has no magnetic dipole moment, we see that the atomic spin density is fully described by a $d_{x^2-y^2}$ form factor. On the bottom panel of Figure \ref{fig:Fig1}e we show the radial profile of the order parameters that are constructed by adding the contributions of the $d_{x^2-y^2}$ ($g_{2}\equiv Y_{4,2}$) partial wave on $A$, $B$ and $O$ sublattices. Note that for this material, the ferroic order parameters are comparable in size with the $s$-wave antiferroic Néel order. In the Supplementary Material, we show the relative size of this atomic spin density form factors, with respect to the $s$-wave magnetic dipole, for several  altermagnetic candidates with a Lieb lattice structure.

 We now focus in the electronic band structure of \kvso\ \cite{Jiang2025}, which has the same spin density structure described before, but in a bulk tetragonal crystal. To understand the spin polarization of the band structure, the following minimal model is instructive \cite{Fernandes2023}:

 \begin{equation}
 	\begin{split}
 		h(\mathbf{k})	=& \lambda_1 (k_x^2-k_y^2)\sigma_z  + \\
 		&  \lambda_2 (k_y-k_x)k_z \sigma_{[110]} +  \lambda_2 (k_x+k_y)k_z \sigma_{[\bar{1}10]}\;,
 		\label{eq:model1}
 	\end{split}
 \end{equation} 
 where $\mathbf{k}$ is the wave-vector and $\sigma_z$, $\sigma_{[110]}$, and $\sigma_{[\bar{1}10]}$ are defined as $\boldsymbol{\sigma}\cdot \mathbf{v}$, with $\mathbf{v}$ a unit vector along $[001]$, $[110]$ and $[\bar{1}10]$ respectively. $\lambda_1$ captures the dominant nonrelativistic $d$-wave spin splitting, and $\lambda_2$ ($\lambda_2<< \lambda_1$) captures additional splittings that appear when SOC is included.

We first focus on the nonrelativistic limit ($\lambda_2=0$ in the previous model). The band structure calculated by DFT is shown in  Fig.~\ref{fig:Fig2}a, with spin splitting reaching a 1 eV scale. The spin polarization of the energy iso-surfaces in momentum space (Fig.~\ref{fig:Fig2}c) shows the same $d$-wave character as the ferroically ordered $d_{x^2-y^2}$ projection of the spin density in direct space (Fig.~\ref{fig:Fig2}b). A key factor behind the large splitting in this family of materials is the Lieb lattice structure. Here, the anisotropy of the exchange enters at the second $A-A$ neighbor. If we consider the $A$ sublattice, the links with neighbors at $(a,0,0)$ and $(0,a,0)$ are different, because the first one contains the O-site in the middle. This anisotropic exchange at close distance can lead to large spin splitting and strength of the altermagnetic order.
 
 When SOC is included in the DFT calculations (Néel order along $[001]$), the previously degenerate nodal planes show now a weak $d$-wave spin-splitting, with spin polarization normal to the respective plane (see terms proportional to $\lambda_2$ in Eq. \ref{eq:model1}). Thus, when looking at the plane in momentum space normal to the $[110]$ axis that crosses the $\Gamma$ point (Fig.~\ref{fig:Fig2}i), the component $S_{110}$ shows a $d$-wave character, with nodal planes normal to the $[\bar{1}10]$ and $[001]$ axes. The corresponding $S_{110}$ component of the direct-space spin density displays the same $d$-wave nodal structure (Fig.~\ref{fig:Fig2}f). Analogous correspondence can be seen for the spin component  $S_{\bar{1}10}$ in Fig.~\ref{fig:Fig2}g and Fig.~\ref{fig:Fig2}j. Even with SOC,  there is a clear correspondence between the nodal structure of the ferroically ordered spin density in direct space, and the spin polarization of the band structure in momentum space. Each spin component displaying a different $d$-wave form-factor, is consistent with the analysis of Ref.~\onlinecite{Fernandes2023}. 
An important consequence of the existence of two nodal planes for each of the three independent spin directions is that the band structure will only be spin-degenerate along nodal lines in the presence of SOC rather than nodal planes.

\textit{Pure atomic altermagnet candidate}|We consider \bcoo, a double perovskite with face-centered cubic unit cell (space group $Fm\bar{3}m$) with Os at the $4a$ Wyckoff site \cite{Thompson2014}. This material attracted significant attention due to unexpected oscillations in zero-field muon spin relaxation  \cite{Thompson2014}, which suggest a phase transition with TRSB at $T\sim 50\;K$. Based on neutron spectroscopy and $X$-ray diffraction measurements, it was proposed that this phase transition can be explained by a ferro-octupolar phase   \cite{Maharaj2020}, with magnetic octupoles ordered on the Os $5d^2$  electrons. Later on, this claim was supported from an \textit{ab initio} perspective  \cite{FioreMosca2022}. To our knowledge, the connection between this octupolar state and the spin splitting in the band structure remains unexplored.

\begin{figure*}[ht!]
	\centering
	\includegraphics[scale=0.21]{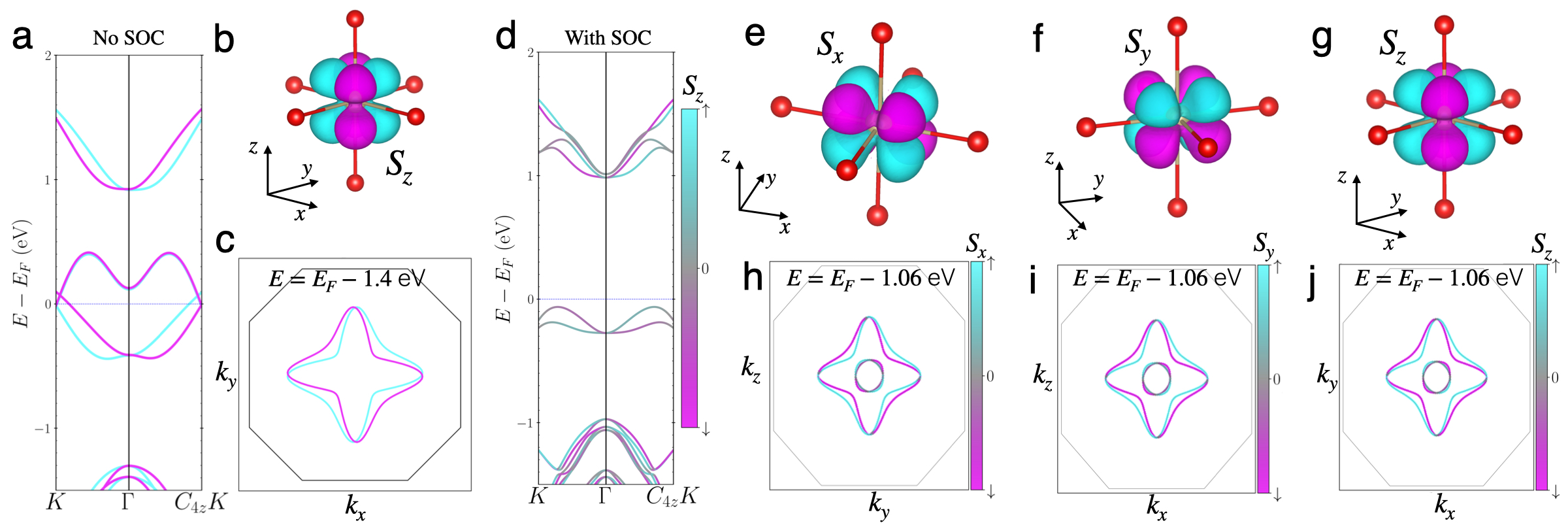}
	\caption{DFT+U results for \bcoo\ without (a–c) and with (d–j) SOC.
		(a,d) Band structures colored by $S_z$ spin polarization.
		(b) Iso-surface of the direct-space spin density (no SOC) around the Os site.
		(c) Energy iso-surface in the $k_z = 0$ plane.
		(e–g) Spin density iso-surfaces with SOC, projected along $[100]$, $[010]$, and $[001]$.
		(h–j) Energy iso-surface cuts in planes normal to $[100]$, $[010]$, and $[001]$, colored by spin polarization normal to each plane.}
	\label{fig:Fig3}
\end{figure*}

Despite strong SOC in \bcoo, the non-relativistic limit remains instructive, as we still find a correspondence between direct- and momentum-space order parameters. We compute the spin density using DFT+U (Liechtenstein scheme) with $U = 3.2$ eV and $J = 0.5$ eV, following the approach of Ref.~\onlinecite{FioreMosca2022}.
Without SOC, the collinear spin density around the Os site is shown in Fig.~\ref{fig:Fig1}c. Its partial wave expansion (Fig.\ref{fig:Fig1}f) reveals $d_{xy}$ and $g_{\bar{2}} \equiv Y_{4,-2}$ components, with nodal planes along $[100]$ and $[010]$.
No magnetic dipole moment is present, identifying \bcoo\ as a pure atomic altermagnet, with spin point group ${}^24/{}^1m{}^2m{}^1m$. The iso-energy surfaces (Fig.\ref{fig:Fig3}c) exhibit the same nodal structure as the spin density in direct space.

In Fig.~\ref{fig:Fig3}a we show the band structure along a path in the $k_z=0$ plane that bisects the two nodal planes. We identify a spin splitting of 0.2 eV just below the Fermi level. We emphasize that, because \bcoo\ is a Mott insulator, bands crossing the Fermi level in Fig.~\ref{fig:Fig3}a will be pushed away by the on-site repulsion. A full treatment of strong correlations is beyond the scope of this work, which focuses on the symmetry-enforced shape of the spin density. In this regard, we note that the $d_{xy}$ form factor of the spin density obtained via DFT+U is consistent with the strong-coupling perspective on the octupolar moment of Os $5d^2$, which can be understood as the $z$-component of the pseudo-spin that characterizes the non-Krammers doublet ground-state of the Os atom, arising from the combined effects of crystal fields and SOC on the localized $J=2$ state  \cite{Paramekanti2020,Voleti2021}.

Upon inclusion of SOC, the spin density in \bcoo\ becomes strongly noncoplanar, with equivalent $d$-wave form factors along $x$, $y$, and $z$, consistent with Refs.~\onlinecite{FioreMosca2022} and \onlinecite{Fernandes2023}. Figure~\ref{fig:Fig3}d shows the band structure (with SOC) along a path in the $x-y$ plane, bisecting the planes normal to  $[100]$ and $[010]$ axes. The $z$-component of the spin polarization is shown in color scale. SOC opens a ~1 eV band gap. We note, however, that strong correlations will make the system a Mott insulator.

In analogy with Eq.~1, a simple model capturing these noncoplanar altermagnetic $d$-wave form factors is given by  \cite{Fernandes2023}:

\begin{equation}
	h(\mathbf{k})	= \lambda k_x k_y \sigma_z  + 
	\lambda k_z k_x \sigma_{y} +  \lambda k_y k_z \sigma_{x} \text{ .}
\end{equation}
Note that, because of the cubic symmetry of the crystal structure, and in contrast to \kvso,  all spin components have the same prefactor $\lambda$.

The DFT spin density projected along $x$, $y$, and $z$ is shown in Figs.~\ref{fig:Fig3}e–g. Below, Figs.~\ref{fig:Fig3}h–j show energy iso-surfaces with spin polarization (component normal to the plane) in color. Comparing the DFT calculations with Eq.~2, we can understand the spin density as 3 orthogonal $d$-wave form factors corresponding to the $x$, $y$ and $z$ spin components. The nodal structure of the spin density is reflected in the spin polarization of the energy iso-surfaces.

The $d$-wave form factors in \bcoo\ arise from an interplay between correlations and orbital order. The on-site occupation matrix of the $d$ orbitals captures the  $d$-wave character \cite{FioreMosca2022}, which then gets imprinted in both the direct-space spin density and the spin splitting in the band structure. A similar orbital-order mechanism was recently studied in a Hubbard square lattice model \cite{Giuli2025}.

\begin{figure}[ht!]
	\centering
	\includegraphics[scale=0.22]{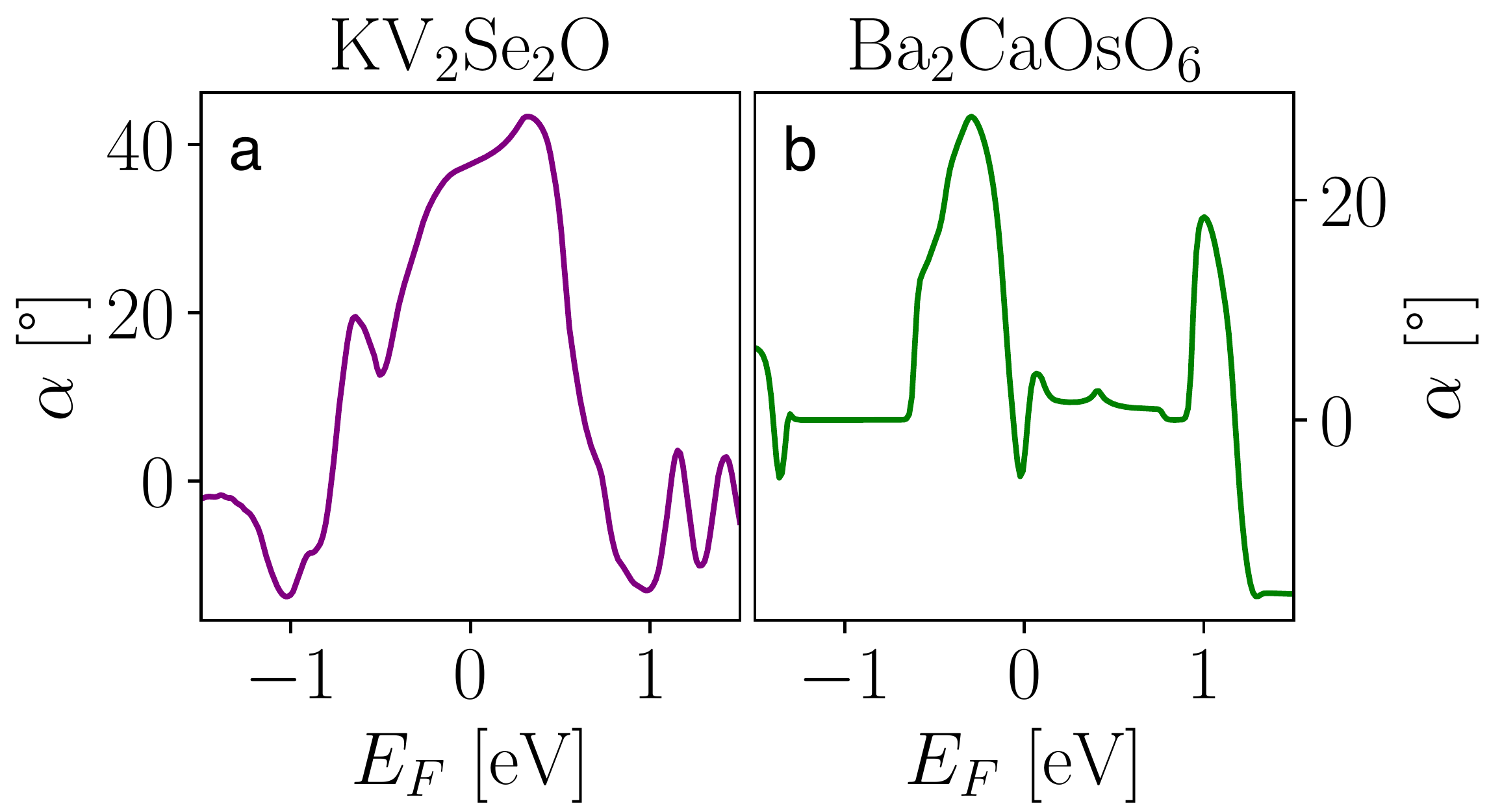}
	\caption{Spin-splitter angle from non-relativistic DFT for \kvso\ (a) and \bcoo\ (b).}
	\label{fig:Fig4}
\end{figure}

\textit{Spin currents} -- We compute spin currents in the non-relativistic limit for \kvso\ and \bcoo. We quantify the spin-splitter angle $\alpha$ \cite{Gonzalez-Hernandez2021} and the GMR coefficient \cite{Smejkal2022GMR}.
Figure~\ref{fig:Fig4} shows $\alpha$ as a function of Fermi energy, peaking at 26$^\circ$ for \bcoo\ and 42$^\circ$ for \kvso. The GMR coefficient in \kvso\ reaches 800\% (see SI), nearly an order of magnitude above original prediction in RuO$_2$  \cite{Smejkal2022GMR}.
Comparable values are expected in related metallic Lieb-lattice altermagnets such as RbV$_2$Se$_2$O \cite{Zhang2025a}.
In the SI, we comment on the nature of the spin conductivity tensor in \bcoo\, when the spin density is described by three noncoplanar $d$-wave form factors.

\textit{Discussion}|While the notion of multipolar magnetism is naturally connected with altermagnetism  \cite{Bhowal2024,Fernandes2023}, one should be careful when comparing these two concepts. Not every direct-space ferroically ordered magnetic multipole, corresponds to the collinear $d$-, $g$-, or $i$-wave altermagnetic order with the characteristic anisotropic nodal spin polarization in the momentum-space electronic structure. As an illustrative example, consider Mn$_3$Sn,
which has a phase between $T_{N2}=50$K and $T_{N1} = 420$ K in which Mn magnetic moments form an inverse triangular spin structure. Although the material exhibits octupolar order $T_x^{\gamma}$
 \cite{Suzuki2017}, its magnetic order on the crystal is non-collinear and, therefore, is not altermagnetic. Correspondingly, the resulting spin polarization in the momentum space  takes noncollinear and nodeless form \cite{Hellenes2023}. 
 

\textit{Conclusions}|~In this work we extend the known mechanisms that induce altermagnetism, to include pure atomic altermagnetism. The interplay between strong correlations and orbital occupation (possibly with strong SOC) generates a spin density which lacks magnetic dipole moments, while having a $d$-wave component which is also reflected in the corresponding spin-polarization symmetry of the electronic structure in the momentum space. We identified pure atomic altermagnetism  with $d$-wave symmetry in the non-relativistic limit of Ba$_2$CaOsO$_6$. When including SOC,  the spin density has the structure of three noncoplanar (and orthogonal) $d$-wave form factors. In all cases, we identify one-to-one correspondence between the ferroically ordered $d$-wave harmonic of the spin density in direct space, and the angular dependence of the spin splitting around the $\Gamma$ point.

In a family of compounds in the Lieb lattice (e.g. \kvso) we demonstrate the coexistence of dipole and higher-order partial-wave components on the V-sites, with the absence of the dipole and the presence of the higher-order partial-wave component (atomic altermagnetism) on the O-sites. We estimate the spin-splitter angle and the GMR coefficient in this material as 42 degrees and 800\%, respectively.

This work thus extends the search for altermagnetic candidates to materials that, while breaking time-reversal symmetry, do not have dipolar order. The detection of the magnetic order without local magnetic dipoles might represent an intriguing challenge for the conventional techniques like neutron scattering. The spin-polarized electronic structure, and the spintronic responses associated with the altermagnetic order, can serve as experimental probes of these higher-order partial-wave forms of magnetic ordering. Besides \bcoo, there is a range of materials that have been previously identified as candidates for magnetic octupolar order that may deserve further investigation, in the context of the higher-partial-wave ordering and atomic altermagnetism  \cite{Santini2000,Lovesey2003,Aoki2008,Matsumura2009,Sibille2020}, including the metallic cage compound PrV$_2$Al$_{20}$, which has recently drawn significant attention \cite{Tsujimoto2014,Freyer2018,Patri2019,Sorensen2021,Ye2024}.

\textit{Acknowledgments}|RJU, VKB, RZ, JS and LS acknowledge funding by the Deutsche Forschungsgemeinschaft (DFG, German Research Foundation)- TRR 173– 268565370 (project A03) and TRR 288– 422213477 (project A09 and B05). TJ acknowledges Ministry of Education of the Czech Republic Grant CZ.02.01.01/00/22008/0004594 and ERC Advanced Grant no. 101095925. R.M.F. was supported by the Air Force Office of Scientific Research under Award No. FA9559-21-1-0423. The authors acknowledge the computing time granted on the supercomputer Mogon at Johannes Gutenberg University Mainz (hpc.uni-mainz.de).

\bibliographystyle{apsrev4-1}

\bibliography{atomic_AM_mendeley}

\end{document}